\def\et{{\rm et al}~}
\documentstyle{mn}
\title{MONTE--CARLO SIMULATIONS OF STAR CLUSTERS I. FIRST RESULTS.}
\author[Mirek Giersz]{Mirek Giersz$\rm ^{1}$ \\
$\rm ^1$ N. Copernicus Astronomical Center, Polish Academy of Sciences, ul.
Bartycka 18, 00--716 Warsaw, Poland\\}

\date{\small Accepted 199- --- --. Received 199- --- --; in original form 1996
December 1 --- ???}

\input epsf.sty

\begin{document}

\maketitle

\begin{abstract}

A revision of Stod\'o\l kiewicz's Monte--Carlo code is used to simulate
evolution of star clusters. The new method treats each {\it superstar}
as a single star and follows the evolution and motion of all individual
stellar objects. The first calculations for isolated, equal--mass $N$--body
systems with three--body energy generation according to Spitzer's
formulae show good agreement with direct $N$--body calculations for
$N=2000$, $4096$ and $10000$ particles. The density, velocity, mass
distributions, energy generation, number of binaries etc. follow the
$N$--body results. Only the number of escapers is slightly too high
compared to $N$--body results and there is no level off anisotropy for
advanced post--collapse evolution of Monte--Carlo models as is seen in
$N$-- body simulations for $N \leq 2000$. For simulations with $N >
10000$ gravothermal oscillations are clearly visible. The calculations of
$N=2000$, $4096$, $10000$, $32000$ and $100000$ models take about $2$,
$6$ $20$, $130$ and $2500$ hours, respectively. The Monte--Carlo code is
at least $10^5$ times faster than the $N$--body one for $N=32768$ with
special--purpose hardware (Makino 1996ab). Thus it becomes possible to
run several different models to improve statistical quality of the data
and run individual models with $N$ as large as $100000$. The
Monte--Carlo scheme can be regarded as a method which lies in the middle
between direct $N$--body and Fokker--Planck models and combines most
advantages of both methods.

\end{abstract}

\vspace{1.0cm}


\section{INTRODUCTION.}

Our knowledge about the stellar content, kinematics, and the
influence of the environment on observational features of globular
clusters and even richer stellar systems are increasing dramatically
(Janes 1991, Djorgovski \& Meylan 1993, Smith \& Brodie 1993,
Hut \& Makino 1996, Meylan \& Heggie 1997).
First, observations are reaching the point where segregation of mass
within globular clusters can be observed directly and quantitatively.
Second, observations have revealed that clusters with dense (collapsed)
cores are relatively more concentrated to the galactic center than
uncollapsed ones. Thus the influences of the environment and mass
spectrum are crucial for cluster evolution. Third, observations give
clear evidence that post--collapse globular clusters have bluer
cores. This suggests strong influence of dynamical interactions
between stars on observational properties of globular clusters. Fourth,
recent observations
show that many different and fascinating types of binaries and binary
remnants are present in abundance in globular clusters. Binaries, in
addition to being a diagnostic of the evolutionary status of clusters,
are directly involved in the physical processes of energy generation,
providing the energy source necessary to stop the core collapse and then
drive the core expansion. So, to model the evolution of real stellar
systems and make meaningful comparison with observation one has to take
into account the complex interactions between stellar evolution,
stellar dynamics and the environment. Of course all these demands can be
fulfilled by direct $N$--body codes (but even the $N$--body method
will have trouble with stellar evolution of binary stars). But they are very
time--consuming and they need
a special--purpose hardware to be run efficiently (Makino 1996ab). Another
possibility is to use a code which is very fast and properly reproduces
the standard relaxation process and at the same time provides a clear
and unambiguous way of introducing all the physical processes which are
important during globular cluster evolution. This task might seem
unachievable, but
actually this kind of code was in use in the past. Monte--Carlo codes,
which use a statistical method of solving the Fokker--Planck equation
provide all the necessary flexibility.  They were developed by Spitzer
(1975, and references therein) and H\'enon (1975, and references
therein) in the early seventies, and substantially improved by Marchant
\& Shapiro (1980, and references therein) and Stod\'o\l kiewicz (1986a,
and references therein).  Unfortunately, lack of fast computers with
sufficient memory at that time and development of the direct Fokker--Planck
and gaseous models contribute to the abandonment of this method. But
recent developments in computer hardware,  speed and memory now make it
possible to run a Monte--Carlo code efficiently, even on general--purpose
workstations. The great advantages of this method,
beside of its simplicity and speed, are connected with the inclusion of
anisotropy and with the fact that added realism does not slow it down. The
Monte--Carlo method can practically cope as easily as the $N$--body method
with internal freedom of single and binary stars and external environment,
with one exception, a stellar system must be spherically symmetric.

The Monte--Carlo code can have another possible use. Despite the
simplified nature of continuum models (Fokker--Planck and gaseous models)
they will continue for a while to be the most commonly used codes for
stellar dynamical evolution. The Monte--Carlo models can be used to
optimise physical free parameters and approximations of continuum models
to check their validity as it was done in comparison between small
$N$--body simulations
and continuum ones (Giersz \& Heggie 1994ab, Giersz \& Spurzem 1994).
This procedure should further increase our confidence in results
obtained by Fokker--Planck or gaseous simulations. On the other hand the
Monte--Carlo techniques can be incorporated in continuum models to describe
the stochastic processes of binary formation, energy generation and
movement (Spurzem \& Giersz 1996, Giersz \& Spurzem 1997). This, for
example, will enable a very
detailed investigation of evolution of primordial binaries in evolving
background given by an anisotropic gaseous model.

The plan of the paper is as follows. In Section 2 a short review of the
`old' and `new' Monte--Carlo methods will be presented. In Section 3
the first results of the `new' Monte--Carlo simulation will be
presented. And finally in Section 4 the conclusions and future
development of the code will be discussed.

\section{MONTE--CARLO METHOD.}

\subsection{Basic ideas.}

The Monte--Carlo method can be regarded as a statistical way of solving
the Fokker--Planck equation. Similarly as the direct Fokker--Planck method
it is based on three main assumptions; i.e. {\bf (1)} the
gravitational field can be divided on two parts: a smooth,
mean field and an irregular, fluctuating field, which causes the
cluster evolution, {\bf (2)} the system evolves due to distant two--body
interactions through a sequence of essentially steady states, {\bf (3)}
the system is spherically symmetric.

The basic idea behind the Monte--Carlo method takes full advantage of
these assumptions. During a time interval $\Delta t$, much smaller
than the relaxation time and larger than the crossing time, the fluctuating
gravitational field can be
neglected in a first approximation and the system can be regarded as
being in a steady state. Because of spherical symmetry of the mean
gravitational field the motion of stars is fully described by simple
analytical formulae, an orbit is a plane rosette confined between
$r_{\rm min}$ and $r_{\rm max}$ radii, which are defined (in a given
potential) by energy $E$ and angular momentum $J$ of a star.
However, the fluctuating field causes slow and random changes of the
orbit parameters, $E$ and $J$. This effect is small over $\Delta t$, but it
builds up
and becomes significant over the relaxation time scale and it has to be
taken into account. To compute it, the influence of all stars in the
system at all positions on a test star orbit during the time
interval $\Delta t$ should be considered. It seems that a direct $N$--body
integration has to be performed to calculate the perturbation. But
instead of doing this the standard Monte--Carlo tricks can be applied.
The perturbation of a test star orbit is a random quantity, so only its
statistical properties matter -- the first and second
order moments. The exact value of each perturbation is unimportant. The
procedure to calculate perturbations is as follows: {\bf (1)} instead of
integrating a sequence of uncorrelated small--angle  perturbations along the
orbit, a single perturbation is computed at a randomly selected point of
the orbit, {\bf (2)} instead of considering the effect of all stars in
the system, the perturbation is computed locally from a
randomly chosen star, {\bf (3)} the computed single perturbation is
multiplied by an appropriate factor in order to account for the
cumulative effect of all small individual encounters with the rest stars
in the system during the past time step. If the procedure is
correctly set up, the evolution of the artificial system will be
statistically the same as the evolution of the real one.

The way of implementing this basic strategy divides Monte--Carlo codes on
three different groups; referred to as `Princeton', `H\'enon' and
`Cornell' methods (Spitzer 1987, and references therein). Briefly, in
the `Princeton' method the stellar orbits are directly integrated with
velocity perturbation {$\Delta\bmath v$} chosen to represent proper
averages over all types of encounters at each orbital position.
{$\Delta\bmath v$} is obtained directly from the standard diffusion
coefficients computed for isotropic and Maxwellian velocity distribution
of the field stars. The direct integration of the star orbits and
computation of velocity changes produced on a single orbit makes it
possible to examine violent relaxation and to investigate the rate of
escape from an isolated system, respectively. The main disadvantages are
that the velocity distributions of test and field stars are different
and that the method requires more computing time than other methods.

In the `H\'enon' method to compute velocity perturbations produced by
encounters  the theory of two--body relaxation is used to integrate over
the impact parameters of all encounters during the time $\Delta t$.
Then the main square cumulative value of deflection angle is computed.
The effective impact parameter is chosen to give the cumulative deflection
angle in a single encounter. A big advantage of this method is that the
velocity distributions of the test and field stars are the same and the
computing time scales with $N$ nearly linearly.

In the `Cornell' method the changes of energy $\Delta E$ and angular
momentum $\Delta J$ resulting from encounters during {\bf integral}
number of orbits are computed by use of five orbit--averaged diffusion
coefficients: $<\Delta E>_{\rm orb}$, $<\Delta J>_{\rm orb}$, $<\Delta
E^2>_{\rm orb}$, $<\Delta J^2>_{\rm orb}$ and $<\Delta E \Delta
J>_{\rm orb}$. In
order to compute these coefficients the velocity distribution of the
field stars is equal to suitable isotropized distribution of the test
stars. This method is especially suitable for investigation of physical
processes which occur on an orbital time--scale, for example such as:
escape of stars or their capture by a central black hole.

Each of these Monte--Carlo implementation was successfully used in
simulations of evolution of globular clusters and galactic nuclei. Now I
would like to proceed to a more detailed description of Stod\'o{\l}kiewicz's
Monte--Carlo scheme, a version of `H\'enon' method, which is the base
of the new Monte--Carlo code presented here.

\subsection{Stod\'o{\l}kiewicz's Monte--Carlo scheme.}

The real power of Monte--Carlo codes was demonstrated by
Stod\'o{\l}kiewicz (1982, 1985, 1986a). He substantially improved
H\'enon's version of Monte--Carlo code by adding an individual time--step
scheme and a special procedure which very much improves the total energy
conservation. His code was used to model the evolution of globular
clusters influenced by the following processes: formation of binaries by
dynamical and tidal interactions, interaction between binaries and field
stars and between binaries themselves, collisions between stars, stellar
evolution, the tidal field of the Galaxy and tidal shocks. They were
unique calculations and have never been repeated or superseded by
anybody.

Because the detailed description of Stod\'o{\l}kiewicz's code was
presented more than ten years ago and since then the method was abandoned,
I will very briefly describe the basic ingredients of the code. More details
can be found in Stod\'o{\l}kiewicz (1982, 1986a).

The evolution of a stellar system is governed by the changes with time
of energy (per unit mass) $E$ and angular momentum (per unit mass) $J$
of all stars in it. According to the assumptions discussed in the previous
section these changes are described by the following equations

\begin{equation}
{{\rm d}E \over {\rm d}t} = {\partial U(r,t) \over {\partial t}} + {\left(
{{\rm d}E \over {\rm d}t} \right)_e},
\end{equation}

\begin{equation}
{{\rm d}J \over {\rm d}t} = {\left( {{\rm d}J \over {\rm d}t} \right)_e},
\end{equation}
were $U(r,t)$ is the gravitational potential at distance $r$ from the
cluster center and time $t$. The first term on right-hand-side of
equation (1)
describes the changes of stellar energies caused by the evolution of the
gravitational potential, i.e. by changes of the mass distribution --
the density in the bulk of system changes slowly while in the core
it grows rapidly. The terms with subscript `e' describe the encounter
effects connected with the relaxation process; the driving mechanism of the
system evolution.

In the Monte--Carlo method the whole system is divided into $K$ {\it
superstars} each consisting of a certain number of stars with the same
mass $m$, distance $r$ from the cluster center, radial $v_{\rm r}$ and
tangential $v_{\rm t}$ velocities. For each {\it superstar} changes of $E$
and $J$  are computed by the procedures which simulate
relaxation processes and changes of gravitational potential. These
simulations should give the correct statistical distributions (in
practice the mean values) of $\Delta E$ and $\Delta J$ for all {\it
superstars}.

The relaxation process of the whole system in the time interval $\Delta
t$ is simulated as a number of single two--body encounters of neighbouring
{\it superstars}, arranged according to their distance from the center.
In order to obtain a good overall effect of the system relaxation over time
$\Delta t$ a very careful approach has to be taken to get the proper
effective deflection angle $\beta$. This is done in the following way.
For a single encounter
of two stars with masses $m_1$ and $m_2$ and velocities ${\bmath v}_1$ and
${\bmath v}_2$ the velocity change of the first star is given by

\begin{equation}
m_1 (\Delta{\bmath{v}}_1)^2 = 4{m_1 m_2^2 \over{(m_1 + m_2)^2}} w^2
{\rm sin}^2{\beta \over 2},
\end{equation}
where $w$ is the relative velocity of interacting
stars, $\beta$ is the angular deflection in the relative orbit of interacting
stars. On the other hand the mean overall result of encounters of a test
star with the field stars during time $\Delta t$ is approximately equal to
(H\'enon 1975)

\begin{equation}
<m_1 (\Delta{\bmath{v}}_1)^2> = 8 \pi G^2 n \Delta t <m_1 m_2^2 w^{-1}>
{\rm ln}(\gamma N),
\end{equation}
where $G$ is the gravitational constant, $n$ is the number
density and ${\rm ln}(\gamma N)$ is the Coulomb
logarithm. Comparison of equations (3) with (4) leads to the following
definition of deflection angle $\beta$

\begin{equation}
{\rm {sin}}^2 {\beta \over 2} = 2 \pi G^2 {(m_1 + m_1)^2 \over w^3} n
\Delta t {\rm ln}(\gamma N).
\end{equation}
Equation (5) connects the relaxation process with the evolutionary time.
This is the only equation in which time appears explicitly. The value of
$\beta$ depends on the length of the time--step. The larger time--step the
larger $\beta$. If the chosen time--step is too large (so that the
right-hand-side of equation (5) exceeds unity) the system is
underrelaxed and computation depends on the length of the time--step.
Therefore, in order to obtain the correct description of the relaxation
process, simulations should be conducted with the time--step sufficiently
small (smaller than the local relaxation time and lager than the local
crossing time) and, as well, dependent on the position in the system (the
local relaxation time changes strongly in the system, increasing towards the
center). This is done by dividing the whole system on a certain number
of zones, for which according to equation (5) the average individual
time--steps are computed with $\beta$ kept in certain boundaries
(between $0.025$ and $0.05$ for equal mass stars). The
boundaries for $\beta$ are chosen experimentally to ensure that the
results of simulations are practically independent on the chosen time--step.
The zones with the same time--step are collected together forming
a larger superzone. The time--steps for successive superzones can only
differ by factor of two. The time interval after which the succeeding
model of the whole cluster is completed is chosen experimentally and is
equal to $0.08$ (about one half of the initial half-mass relaxation
time). This procedure allows to compute encounters in
each part of the system with the time--step several times shorter than
the local relaxation time. At the end of the encounter procedure (in the
considered superzone) the new velocities for each two interacting stars are
computed using the standard scheme (H\'enon 1971). This scheme takes
into consideration that the plane of relative motion of interacting
stars and their relative orbit in this plane can be oriented randomly.
This concludes the relaxation step.

Now, the new positions of all {\it superstars} in the actually computed
superzones are chosen. All {\it superstars} with positive total energy
or with distance greater than the tidal radius (for non--isolated system)
are treated as  escapers. New positions of remaining stars are selected
randomly in their orbits between the pericentre and a maximum distance
with the probability proportional to the time, which the star spends in a
given place in the orbit. The maximum distance is chosen to be the
smallest distance of either apocentre or the position of the last {\it
superstar} in the actually computed superzones. As a result of this
procedure, when many successive encounters are included, the $\Delta E$
and $\Delta J$ are correctly averaged over the test star orbit. After
evaluating the new positions (for all {\it superstars} in the actually
computed superzones) a new distribution of {\it superstars} has been
obtained. This leads to small changes of the mass distribution. The
resulting changes of potential with time induce changes in
mechanical energy of the {\it superstars}. This energy change for the
$i$-th {\it superstar} in time $\Delta t$ is given by

\begin{equation}
\Delta E_{\rm i} = \int {\partial U(r,t) \over \partial t}{\rm d}t,
\end{equation}
where the integral is taken along the trajectory of the $i$-th {\it
superstar}. Two points of the trajectory are distinguished; the old
$r_{\rm io}$ and new $r_{\rm in}$ positions. Both, are chosen randomly
in the orbit, which in meantime, changed only slightly due to relaxation
process. So they can be treated as representative for the orbit and
equation (6) can be approximated by the following expression

\begin{equation}
\Delta E_{\rm i} = {1 \over 2} [\Delta U(r_{\rm io}) + \Delta U(r_{\rm in})],
\end{equation}
where $\Delta U = U_{\rm n} - U_{\rm o}$ is the difference between the new
and old
values of the potential in a given point. Comparing equations (6) and
(7) and substituting for $\Delta E_{\rm i}$ the difference between the
new and
old total energies of the {\it i}-th {\it superstar} the new value of
velocity $v_{\rm in}$ is obtained.

\begin{equation}
v_{\rm in}^2 = v_{\rm io}^2 + U_{\rm o}(r_{\rm io}) + U_{\rm n}(r_{\rm
io}) - U_{\rm o}(r_{\rm in}) - U_{\rm n}(r_{\rm in}),
\end{equation}
where $v_{\rm io}$ is the old velocity of the {\it i}-th {\it superstar}
(the old velocity means after relaxation and in old potential).
The new tangential velocity $v_{\rm int}$ is computed using the law of
angular momentum conservation. The radial velocity is equal to $v_{\rm inr}
= \sqrt{v_{\rm in}^2 - v_{\rm int}^2}$. The use of equation (7) to evaluate
the new
velocities ensures that the total energy of the system practically does not
change during the simulations. However, there is an inconsistency in this
procedure. New radial velocities are computed at the end of the time--step,
while new positions of {\it superstars} are selected in orbits
determined by the old potential. Sometimes this leads to difficulties
with determination of new radial velocity. In such situation $v_{\rm inr}$
is set to zero and the missing energy is subtracted from the energy of
next {\it superstar}. The whole cycle: relaxation process, determination
of new positions and determination of new velocities is then repeated for all
superzones.

\bigskip

{\subsection{New implementation of Stod\'o{\l}kiewicz's Monte--Carlo method}

The Monte--Carlo method briefly described in the previous section is not
suitable to correctly represent the very center of the system. In the
core, as a result of the collapse, the density in a small and nearly
uniform region reaches high values. This area is represented by only a
few {\it superstars}. Therefore the statistical properties of this region are
very poorly described. Moreover, {\it superstars} which belongs to the
core take part in processes which are responsible for energy generation
and creation of many different and fascinating types of binaries, binary
remnants and coalesced stars in direct stellar interactions. In order to
properly describe this region and these processes, in the new code
(written from the scratch) each {\it superstar} is treated as a single
star and evolution and motion of all individual objects are followed
(this is done not only in the core but throughout the system). This
improvement is
possible only due to an enormous increase of speed and memory in present
day general--purpose computers. Note, that the individual treatment of all
objects in the system enables, for example, to investigate the influence of
primordial binaries on the system evolution. In Stod\'o{\l}kiewicz's
method all binaries or coalesced stars take part only in relaxation
process. They were neglected in the computation of the gravitational
potential, so the process of mass segregation of binaries and coalesced
stars relative to single stars was not properly described.

Stod\'o{\l}kiewicz's procedure to deal with the problems of radial velocity
determination after the system adjustment (changes of mechanical energy of
the stars due to changes of potential -- equations 6--8 above) was slightly
changed in the present implementation. If for any star $v_{\rm in}^2$ is
smaller than zero the new radial
velocity of a star is set to zero and tangential velocity is computed
according to the angular momentum conservation law. The missing kinetic
energy is accumulated for all stars which fulfil the above criterion.
At the end of relaxation cycle (for the presently computed superzones)
the kinetic energy for each star in the system is decreased by a factor
equal to the ratio of the total kinetic energy reduced by accumulated
missing kinetic energy and the total kinetic energy. This factor is very
close to one. If for any star $v_{\rm inr}^2$ is smaller than zero
and $v_{\rm in}^2$ is bigger than zero the new radial velocity of a star is
set to zero and the new tangential velocity is set to $v_{\rm in}$. The
problem with the determination of the new radial velocity usually occurs
when a star is close to the pericentre or apocentre of its orbit. So the
assumption that the new radial velocity is equal to zero is well
justified. The fraction of `bad' cases is about $0.01$ per cent of all
relaxation events. The total accumulated energy of `bad' cases is only
a few percent of the initial total energy of the system.

As it was stated in the previous section the deflection angle $\beta$
for two interacting stars is chosen to mimic the overall relaxation of
these stars with the rest of the system over the time $\Delta t$. Then
$\beta$ is the accumulated deflection angle and it is usually bigger
than the deflection angle for an individual small--angle interaction. This
can lead to an overestimation of the number of stars which escape from the
system (particularly for stars on very elongated orbits with binding
energy very close to zero). Indeed, preliminary, test Monte--Carlo
simulations showed too high an escape rate comparable to direct $N$--body
results. It is worth to note that the Stod\'o{\l}kiewicz's Monte-Carlo code
(Stod\'o{\l}kiewicz 1982) also gave a too high escape rate compared to
results of $N$--body simulations available at that time.
H\'enon (1961) pointed out that the escape process
is not a diffusive one, but has to be regarded as a two--body interaction
which directly leads to escape of a star. This point was further modified
by Spitzer \& Shapiro (1972), who pointed out that the distribution
function of stars itself evolves on a relaxation time--scale, and then a
star that has diffused to energies a little below the escape limit can
escape in a single two--body encounter in the core. Giersz \& Heggie
(1984a) further modified this point showing that anisotropy has a large
effect on the escape rate. The larger anisotropy the larger escape rate.
So to properly describe the escape process in the Monte--Carlo code
(according to the discussion above) the following procedure was introduced
{\bf only} for stars which escape due to standard relaxation process.
The probability that the closest encounter has impact parameter less
then $p$ in time $\Delta t$ is given by

\begin{equation}
F(x) = 1 - {\rm e}^{-\lambda x}
\end{equation}
where $x=\pi p^2$ is the area of a disc with radius $p$ and
$\lambda = nw \Delta t$ is the number of interactions per unit area. So
the resulting distribution function (probability density) of impact
parameters is as follows

\begin{equation}
f_p(p) = \left({{\rm d}F\over {\rm d}x} \right) \left({{\rm d}x \over
{\rm d}p} \right) = 2 \pi nwp \Delta t {\rm e}^{-\pi nw \Delta t p^2}
\end{equation}
Using the values of $n$ and $w$ from the computation of the relaxation
process (for two considered stars), a new impact parameter is picked up
randomly according to equation (10). The deflection angle is connected with
the impact parameter by the following formula

\begin{equation}
{\rm sin}^2{\beta \over 2} = {1 \over\displaystyle \left(1 + {\strut \left({p\over\displaystyle p_{\rm
o}}\right)^2}\right)}
\end{equation}
where $p_{\rm o} = G(m_1 + m_2)/ w^2$ is the impact parameter for the
$90^{\circ}$ deflection angle. So using again equation (3) and the
scheme described by H\'enon (1971) the new velocities of two interacting
stars can be found. If one of the two stars has positive binding energy
it is regarded as a escaper and second star is kept in the system with
the new velocity, otherwise two stars are kept in the system with the
newly obtained velocity. If no star would escape in the normal
Monte--Carlo step, then it would not escape in the new procedure (the
new procedure is only invoked if escape occurs in the usual procedure).
In principle, therefore, the new procedure underestimates the escape.
However, the underestimate should be small, because it will almost always
be true that the right-hand-side of equation (11) is much smaller than
that of equation (5). The results of Monte-Carlo simulations presented in
the next section strongly suggest that the procedure discussed above 
represents in a proper way physics behind the escape process. Nevertheless,
other explanations of the too high escape rate in Monte-Carlo simulations
comparable to $N$--body simulations are possible. For example, small
deviations from spherical symmetry of the system can cause small changes of
angular momenta of the stars on very elongated orbits (Rauch and Tremaine
1996) leading to fewer escapers in $N$--body simulations. However, this
process can not be investigated by the Monte--Carlo code.

Basically, the improvements mentioned above are the only major
changes to Stod\'o{\l}kiewicz's original code. Other changes are
rather cosmetic and do not have any influence on the code flow or
implementation of any physical processes. However, before
proceeding further I will briefly describe how binaries are
introduced to the code (in the case of single--mass system -- the
case of multi--mass system will be discussed in the next paper).

As it was mentioned at the beginning of this section all stellar objects,
including binaries, are treated (in the new code) as single
{\it superstars}. This allows,
in a simple and accurate way, to introduce to the code the processes of
stochastic formation of binaries and their subsequent stochastic
interaction with field stars and other binaries. The whole procedure is
introduced in a few separate steps. First of all a new binary has to be
formed. The standard formula for the rate of three-body binary
formation (Hut 1985) gives the probability of binary formation as
\begin{equation}
P_{3b} = \int\int 0.9G^5m^5n^3\sigma^{-9}dVdt,
\end{equation}
where $m$ is the mass of single stars and $\sigma$ is the one-dimensional local
velocity dispersion. The integration is over volume and time. The probability
of binary formation is computed for each time step and for each zone containing
three successive {\it superstars}, starting from the zone closest to the system
center. The computed probability is compared with a random number drawn
from the uniform distribution. If the probability is smaller than the
random number a binary is formed from the first and second star in the zone,
at the position of the center of mass of these two stars.
The procedure is repeated for all (three--stars) zones in the system.
For the binding energy of the newly formed binary the value of $3kT$ is
adopted, which is usually used as a minimum binding energy of
permanent binaries (Hut 1985). Before reaching this threshold
energy the binary has been living for some time in the cluster and
interacting with field stars. Assuming that its centre
of mass is in energy equipartition with field stars, the orbit of the
new binary in the cluster can be computed.

A binary living in the cluster is influenced by close and wide interactions.
Wide interactions only change its movement in the system (relaxation
process) and close interactions change also its binding energy.
To simulate close interactions the procedure suggested by Stod{\'o}{\l}kiewicz
(1986a) was adopted. First, the check is performed, whether a close
interaction is due. The probability of binary field star
interaction is computed using Spitzer's formula (Spitzer 1987).
This probability is as follows
\begin{equation}
P_{3b*} = \int \int {5\pi A_SG^2m^3n\over{6\sigma E_b}}dVdt,
\end{equation}
where $E_b$ is the binary binding energy and $A_S$ is a coefficient equal
to $21$. The integration is over volume and time.
This probability is computed for each binary and compared with a
random number drawn from the uniform distribution. If the binary is due to
interaction with a field star the change of its binding energy is computed
according to the distribution function of energy changes (Spitzer 1987),
$f(z)=\cos^6z$, where $z = \arctan(\Delta E_B/E_B)$. The energy,
$\Delta E_B$, generated in
the interaction is distributed between the single star ($2\Delta E_B/3$) and
the binary centre of mass ($\Delta E_B/3$). Knowing the orbit of the binary,
we know its radial and tangential velocities at any point in the orbit.
The absolute value of the recoil velocity of the binary $\Delta v$ is
determined  from the quadratic equation
\begin{equation}
\Delta v^2 + 2(v_r\cos\theta + v_t\sin\theta \cos\phi)\Delta v
   -{{2\Delta E_b}\over m_b} = 0,
\end{equation}
where $v_r$ and $v_t$ are the initial radial and tangential velocities
of the binary centre of mass, respectively, $\theta$ and $\phi$ are the
randomly chosen direction of the recoil determined according to the
distribution $f(\theta) = \sin\theta/2$ for $\theta\,\epsilon\,(0,\pi]$ and
uniform distribution for $\phi\,\epsilon\,(0,\pi]$, respectively. $m_b$ is
the binary mass. The new components of the binary centre of mass velocity
are directly computed from the old velocities, $\Delta v^2$ and chosen
$\theta$ and $\phi$. Having new velocities the new orbit of the binary
is determined. The same procedure is adopted for the determination of
the new orbit and the new radial and tangential velocities of the single
star which just interacted with the binary.

Now the first, pilot results of the evolution of isolated
single mass systems conducted by the new code will be presented.

\section{FIRST RESULTS}

The Monte--Carlo code contains several free parameters, which have to be
adjusted in order to get the proper representation of the system evolution.
The most important parameters are: the boundaries for the deflection
angle $\beta$, the number of stars used to estimate the local density,
the time interval after which the succeeding models of the whole system
are computed, the coefficient in the Coulomb logarithm  and scattering
cross--section for interaction between binaries and field stars. Note:
except the last two parameters, they are not physical but technical,
mainly used to facilitate simulations. The best way of adjusting them
is to compare the new results of Monte--Carlo simulations with direct
$N$--body data. The same strategy was used to optimise the free physical
parameters of the continuum models (Giersz \& Heggie 1994ab, Giersz \&
Spurzem 1994).

High quality statistical data for single--mass $N$--body simulations
are available only for $N = 250$, $500$, $1000$, $2000$, $4096$ and $10000$
(Giersz \& Heggie 1994ab, Giersz \& Spurzem 1994, Spurzem \& Aarseth
1996). The simulations with $N = 250$ and $500$ should be
excluded from the comparison, because strong two--body encounters have
too much influence on  the dynamical evolution of these systems (Giersz
\& Heggie 1994ab, Giersz \& Spurzem 1994). This is in contradiction to
one of the main assumption on which the Monte--Carlo method is based.
Models with $N = 1000$ are extended only up to one collapse time after
the core bounce. This is too small to properly compare the long--term
post--collapse evolution. The same situation is for model $N = 10000$
which is extended just over the time of core bounce. Simulations with $N
= 2000$ and $4096$ are the best for our purposes. They cover evolution
up to twelve collapses time and consist of several separate runs (for
$N=2000$).
Additionally, because of a good statistical quality of the data, model
with $N = 10000$ was used to compare the core collapse and core bounce
phases of the evolution. Only results for $N = 2000$ were averaged over
$25$ simulations, each having the same initial parameters but with
a different sequence of random numbers used to initialise the positions
and velocities of the stars (Giersz \& Heggie 1994a). For other $N$ only
individual simulations were used to compare with $N$--body runs.

For all simulations (of isolated single--mass systems of point mass
particles) presented in this paper the Plummer model was used as the initial
condition. The standard $N$--body units (Heggie \& Mathieu 1986); total
mass $M = 1$, $G = 1$ and initial energy equal to $-1/4$ have been
adopted for all runs. Monte--Carlo time means the $N$--body time divided
by $N/{\rm ln}(\gamma N)$.

Very extensive and time consuming calculations were performed to adjust
the free parameters of the Monte--Carlo code. It should not be surprised
that the best choice of the free
parameters is similar to that chosen by Stod\'o{\l}kiewicz (1982). The
boundaries  for $\beta$ are $0.025$ and $0.05$ ($0.0125$ and $0.025$ for
large $N$). The time interval, between consecutive, complete models, is about
$0.0075$ -- $0.01$ (about one twentieth of the initial half--mass
relaxation time). A lot of care was taken to properly estimate the local
density, which play an important role in determination: the deflection
angle, the number of created three--body binaries and the number of
interactions between binaries and  field stars. Again, the number of
star chosen to determine it was similar to that used by
Stod{\'o}{\l}kiewicz (1982). Results presented by Giersz \& Heggie
(1994ab) suggest that value of $\gamma = 0.11$ for the coefficient in the
Coulomb logarithm and the scattering cross--section given by Spitzer
(1987) assure the best agreement with $N$--body data. Therefore these
parameters were used in Monte--Carlo simulations presented in this paper.
\begin{figure}
\epsfverbosetrue
\begin{center}
\leavevmode
\epsfxsize=80mm \epsfysize=70mm \epsfbox{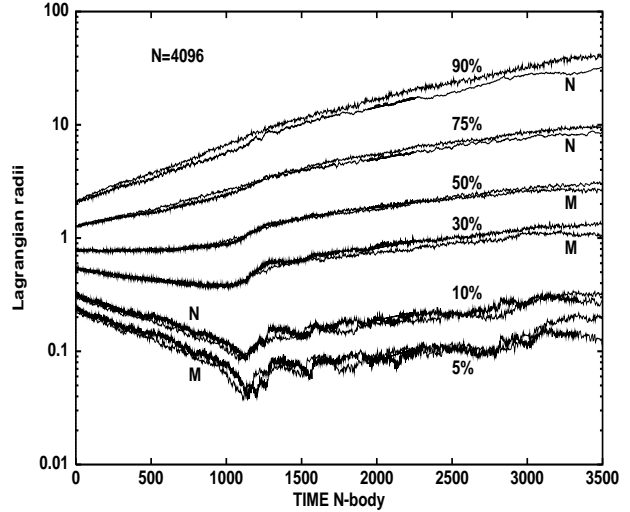}
\end{center}
\caption{ Evolution of Lagrangian radii for $N = 4096$ for Monte--Carlo
model (M) and $N$--body model (N).} \end{figure}

Generally, for both models (Monte Carlo and $N$--body) the phase of core
collapse  is remarkably
similar (Figure 1 for $N=4096$). For $N = 2000$ (not shown here -- see
Giersz 1996), the first
differences start to build up around the time of core bounce. This is
particularly well visible for middle and outer Lagrangian radii. The
rate of system expansion in the $N$--body models is slightly faster
than in the Monte--Carlo models. This behaviour is even more pronounced for
the anisotropy distribution (not shown here -- see Giersz 1996). In the
advanced phase of
core expansion there is no level off anisotropy, feature which is so
characteristic for $N$--body simulations with $N \leq 2000$ (Giersz \&
Heggie 1994b and Giersz \& Spurzem 1994). Probably, for simulations with
small $N$ the relaxation process is slightly overestimated. Encounters
between stars are stronger and more stars are put on elongated orbits
what leads to bigger anisotropy and faster expansion rate. It seems that
this behaviour of Monte--Carlo simulations (for small $N$) can not be
cured by an appropriate selection of the free parameters. For $N =
4096$ in the first phase of the core expansion the agreement between
both models is very good (see Figure 1) but later on the disagreement
starts to build up. Outer parts of the Monte--Carlo models expand
too fast comparably to $N$--body models. And even the inner Lagrangian
radii (except $5$ and $10$ per cent Lagrangian radii) show disagreement
(Monte--Carlo
models evolve too fast), but at least this can be partially associated
with the individual statistical `noise' of simulations. Different
amounts of energy generated by binaries and different times at which
energy generation take place can lead to substantial differences between
simulations for advance post--collapse evolution.

It is worth to note that the collapse time for Monte--Carlo and $N$--body
models is practically the same (taking into account the statistical
spread between simulation with the same $N$). This further confirms the
value of $\gamma = 0.11$, in the Coulomb logarithm, obtained by
comparison of small $N$--body and continuum models ( Giersz \&
Heggie 1994a ).

The energy conserved during the simulations can be divided on the following
parts: total external energy of the system (sum of the total kinetic
and potential energies), total energy of star escapers, total energy of
binary escapers, total internal binding energy of binary escapers and
total internal binding energy of bound binaries. Figure 2 shows for both
models ($N = 4096$) the binding energy of binaries bound to the system.
\begin{figure}
\epsfverbosetrue
\begin{center}
\leavevmode
\epsfxsize=80mm \epsfysize=70mm \epsfbox{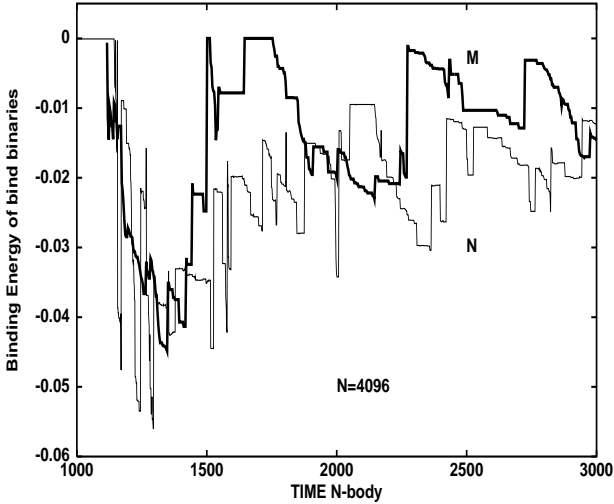}
\end{center}
\caption{The binding energy of binaries bound to the system for $N = 4096$
for Monte--Carlo model (M -- thick line) and $N$--body model (N).}
\end{figure}
The agreement between the models is very good (taking into account the
statistical `noise'). Even the size and amplitude of the energy bump
around the time of core collapse is very similar. This guarantees that
for both models around the time of core bounce the expansion rate and
the shape of the inner Lagrangian radii curves are nearly the same. Later
on, however, the total binding energy of bound binaries is systematically
smaller than in the $N$--body simulations.
\begin{figure}
\epsfverbosetrue
\begin{center}
\leavevmode
\epsfxsize=80mm \epsfysize=70mm \epsfbox{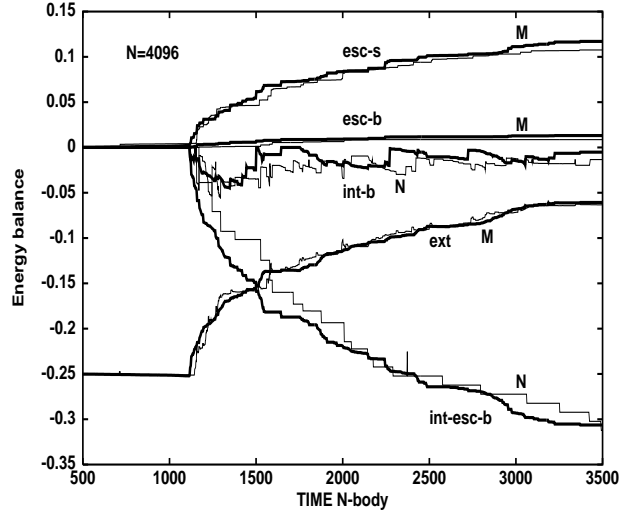}
\end{center}
\caption{Energy balance for $N = 4096$ for Monte--Carlo model (M -- thick
line) and $N$--body model (N). Esc-s - total energy of star escapers,
esc-b - total energy of binary escapers, int-b - total binding energy of
bound binaries, int-esc-b - total binding energy of escaped binaries,
ext - external energy of the system.}
\end{figure}
Correspondingly the total binding energy of escaped binaries is bigger
for Monte--Carlo models to balance the differences for binding energies
of bound binaries. The external binding energy of the system and escape
energies of single and binary stars are remarkably similar (see Figure 3).
For $N = 2000$ agreement between models is not so good (see Giersz 1996).
Around the time of core bounce the bump in the total binding energy of
bound binaries
is smaller and more shallower for Monte--Carlo models than for $N$--body
ones. So the resulting expansion of the middle and outer parts of the
system is less abrupt for Monte--Carlo simulations. Also energy of
escapers and total binding energy of escaped binaries is bigger for
Monte--Carlo models during advanced  post--collapse evolution. This at
least partially leads to higher anisotropy and expansion rate for
Monte--Carlo models for middle and outer Lagrangian radii.

The example of an anisotropy evolution for Monte--Carlo simulations
($N=32000$) is presented on Figure 4.
Anisotropy is defined as $A = 2 - V_t^2/V_r^2$, where $V_t$ and $V_r$ are
the tangential and radial velocity dispersions, respectively. Each was
computed as a mass--weighted average taken over all stars within shells
bounded by consecutive Lagrangian radii. There is no discernible anisotropy
in the innermost shells, up to the Lagrangian radius 10 per cent of the mass.
For intermediate and outermost shells anisotropy starts to increase from the
very beginning, the faster the further from the centre of the system.
For intermediate shells, at the time of core bounce (when the binaries
start to influence the core evolution) sharp increases of anisotropy can be
seen. Then the rate of increase of anisotropy slows down, but there is no
level off anisotropy (Giersz \& Heggie 1994b). Finally, for the outermost
shells, shortly after the core bounce, anisotropy reaches a maximum value,
which is very close to $2$. The outermost parts of the
system are practically populated by stars on radial orbits.
\begin{figure}
\epsfverbosetrue
\begin{center}
\leavevmode
\epsfxsize=80mm \epsfysize=70mm \epsfbox{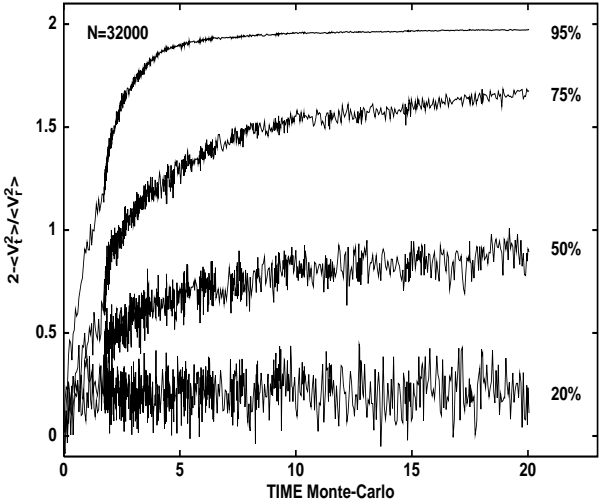}
\end{center}
\caption{Evolution of the anisotropy averaged within mass shells defined by the
Lagrangian radii (15-20, 45-50,70-75, 90-95 per cent) as a function of time
for $N=32000$ for Monte-Carlo simulations.}
\end{figure}

In Figure 5 the number of escapers for $N=4096$ for $N$--body simulations
and
Monte--Carlo simulations with different treatment of escapers (see previous
section) is shown.
\begin{figure}
\epsfverbosetrue
\begin{center}
\leavevmode
\epsfxsize=80mm \epsfysize=70mm \epsfbox{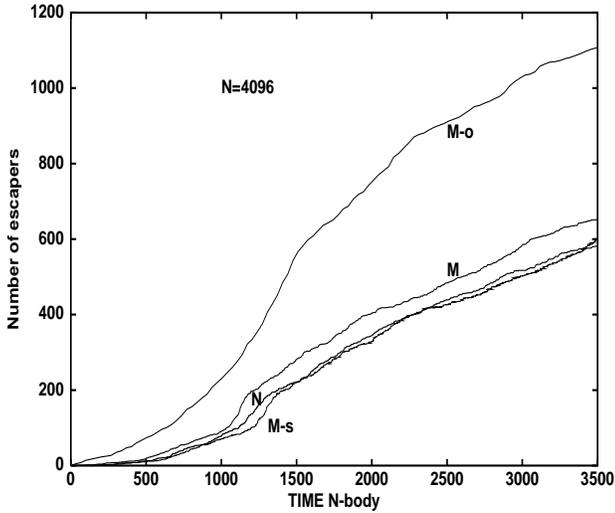}
\end{center}
\caption{Number of escapers for $N = 4096$ for $N$--body model (N) and
Monte--Carlo models: with better treatment of escapers (M), without that
treatment (M-o) and with shifted data (M-s) - see the text for more
explanations}
\end{figure}
Only because of a better treatment of escaped stars (described in the
previous section) the agreement between both curves (labelled by $N$ and
$M$) is reasonably good. Without this procedure differences would be
roughly $80$ per cent (curve labelled by $M$-$o$). For $N$--body
simulations there is a delay in removing escaped stars from the system
(stars have to travel distance up to ten times the half--mass radius -- for
Monte--Carlo simulations are removed instantaneously). To
account for that the Monte--Carlo curve in Figure 5 was shifted in
time by $150r_{\rm h}/r_{\rm ho}$ $N$--body time units (in $N$--body model
the first star escapes at time $155$). The ratio of present to initial
half--mass radii takes in a very simplifying way into account the fact
that in the course of evolution the system expands and more time is
needed for stars to escape. The both curves (labelled by $N$ and $M$--$s$
on Figure 5) come to very close agreement. This confirms the simple idea
about time--dependent shift and source of disagreement between curves
$M$ and $N$ in Figure 5. It is worth to note that the escape rate
for Monte-Carlo simulations (with  'special' treatment of escapers) for
$N=2000$ and $N=10000$  is also in good agreement with $N$--body
simulations.

Comparison between the Monte--Carlo simulations and the $N$--body ones for
$N = 10000$ shows, basically, the same features as in the case of $N =
4096$.

Gravothermal oscillations are the most pronounce feature of the
post--collapse evolution of $N$--body systems with a number of stars greater
than a few thousands. They were observed in gas (Bettwieser \& Sugimoto
1984, Goodman 1987, Heggie \& Ramamani 1989, Spurzem 1994), Fokker--Planck
(Cohn \et
1986, Cohn, Hut \& Wise 1989, Gao \et 1991, Takahashi \& Inagaki 1991)
and recently in $N$--body simulations (Makino 1996ab). The lowest value of
$N$ for which gravothermal oscillations begin to show up is uncertain.
For continuum models it is around $7000$. A pilot $N$--body
simulation of system consisting of $16384$
particles (Makino 1996ab) shows clear oscillations. There are also some
signs of gravothermal oscillations in $N$--body simulations for $5000$
particles (Heggie 1995) and even smaller $N$ (Makino, Tanekusa \&
Sugimoto 1986, Makino 1989). All these results suggest that gravothermal
oscillations should be as well present in Monte--Carlo simulations
(discussed here), at least for $N = 32000$ or more. It is worth to note
that in unpublished Monte--Carlo simulations for $N=100000$ conducted by
Stod{\'o}{\l}kiewicz (1986b) there are some sings of gravothermal
oscillations (see Giersz 1996). In Figure 6 the
evolution of the logarithm of the central density is presented for $N =
4096$, $10000$, $32000$ and $100000$, respectively.
\begin{figure}
\epsfverbosetrue
\begin{center}
\leavevmode
\epsfxsize=80mm \epsfysize=70mm \epsfbox{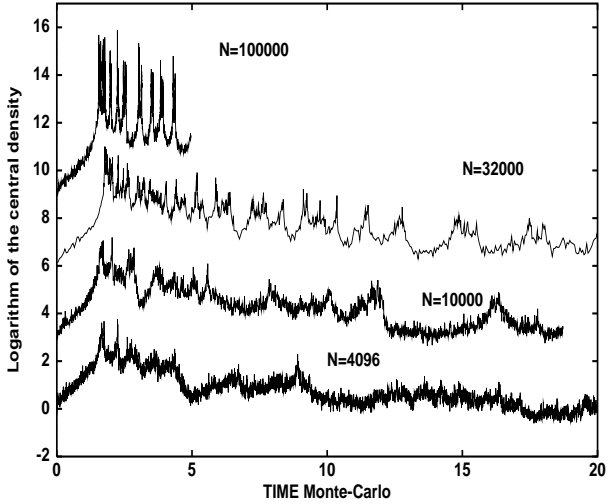}
\end{center}
\caption{Evolution of the central density for $N = 4096$, $10000$,
$32000$ and $100000$. Data are shifted in the logarithm by $3$, $6$ and
$9$ for $N = 10000$, $32000$ and $100000$, respectively.}
\end{figure}
For $N = 4096$ there are some oscillation features, but they are
practically undistinguishable from fluctuations. For $N = 10000$
oscillations are more visible (for example around the time $3$ or $12$), but
still it is difficult to judge if
they are clearly gravothermal. However, for larger values of $N$ there
is no doubt that the gravothermal oscillations are present, as will be
proven in more details below. It should be noted that for oscillations
observed in Monte--Carlo simulations there is no clear transition from
regular oscillations to chaotic ones or from stable expansion to
oscillations. Features observed in gas and Fokker--Planck models (Heggie
\& Ramamani 1989, Breeden \et 1994). However, the present results are
consistent with results obtained by Takahashi \& Inagaki (1991) for
stochastic Fokker--Planck model (stochastic binary formation and energy
generation), by Makino (1996ab) for $N$--body simulations and by Giersz
\& Spurzem (1997) for anisotropic gaseous model with fully
self-consistent Monte-Carlo treatment of binary population.
This further supports the suggestion given by Takahashi \&
Inagaki (1989) that for stochastic systems gravothermal oscillations
are more chaotic and unstable.
\begin{figure}
\epsfverbosetrue
\begin{center}
\leavevmode
\epsfxsize=80mm \epsfysize=70mm \epsfbox{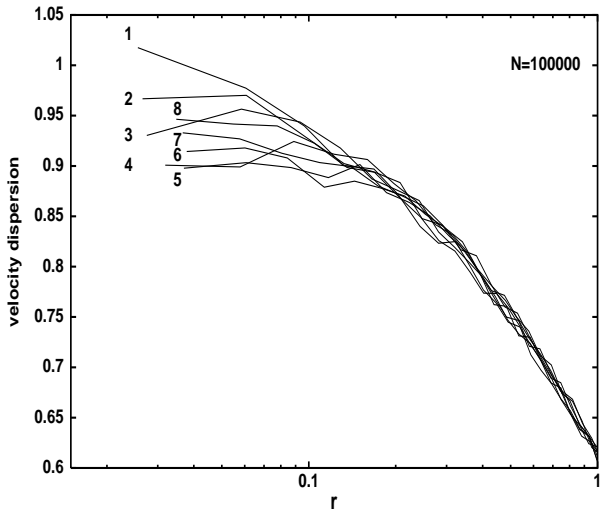}
\end{center}
\caption{Velocity distribution for $N=100000$ for a sequence of time
outputs during the expansion phase around time $2.5$ (see Figure 8). Labels
from $1$ to $8$ are for increasing time.}
\end{figure}
\begin{figure}
\epsfverbosetrue
\begin{center}
\leavevmode
\epsfxsize=80mm \epsfysize=70mm \epsfbox{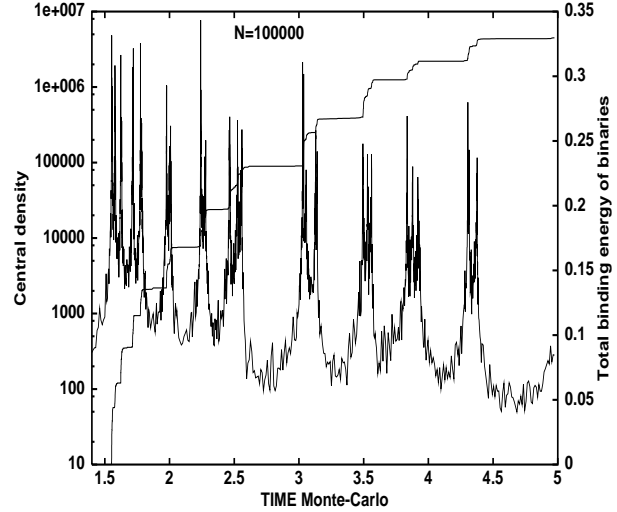}
\end{center}
\caption{Evolution of the central density and the total binding energy
of binaries (stepped curve) for $N = 100000$.}
\end{figure}

Now, lets take a closer look on run with $N = 100000$ particles and examine in
more details the phases of expansions. It is widely accepted that a long
expansion phase without significant energy generation and with temperature
inversion (during this phase) are the most important signatures of
gravothermal expansion (Bettwieser \& Sugimoto 1984, McMillan \& Engle
1996). In Figure 7 the velocity distribution is presented for a sequence
of time outputs during the huge oscillation around the time $2.5$ (see
Figure 8).
Despite the fact that the velocity distribution is plotted every $2$ per cent
Lagrangian radii the temperature inversion is very clearly visible for
curves labelled $2$, $3$ and $4$ (see Figure 7). The temperature inversion is
bigger than $4$ per cent (taking into account that the central
temperature is
slightly smaller than temperature at the $2$ per cent Lagrangian radius).
This
value is in a very good agreement with results of gas (Bettwieser \&
Sugimoto 1984, Heggie \& Ramamani 1989), Fokker--Planck (Cohn et al. 1989)
and $N$--body (Makino 1996ab) models.
Now lets concentrate on the phases of energy generation. Figure 8
shows an enlarged view of the time variation of the central density and
the total binding energy of binaries. It is clear that energy is mostly
generated when system is found in maximum density phases and expansion
is mainly driven without binary energy generation. There are visible
several such expansions. All these expansions continue much more than
hundreds of the central relaxation times. For expansions driven by
binaries it could not continue for more than one hundred central
relaxation time (Makino 1996b). Two expansion phases are exceptional
(around the time $2.5$ and $4.3$).
\begin{figure}
\epsfverbosetrue
\begin{center}
\leavevmode
\epsfxsize=80mm \epsfysize=70mm \epsfbox{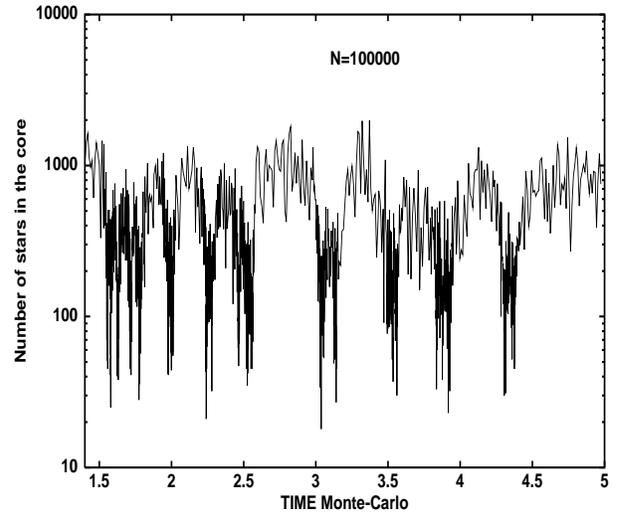}
\end{center}
\caption{Evolution of the number of particles in the core for $N = 100000$.}
\end{figure}
Their duration is much longer than for any other oscillation. There is no
obvious other explanation for that than gravothermal oscillations.
Indeed, results of anisotropic gaseous simulations for $N=100000$ conducted
by Spurzem (1994) show that oscillations presented in  $\zeta$, ${\rm
log}(\rho_{\rm c})$ and  ${\rm log}(v_{\rm c})$ space (where $\zeta =
t_{\rm rc} {\rm d}{\rm ln}(\rho_{\rm c})/{\rm d}t$, $\rho_{\rm c}$,
$v_{\rm c}$ and $t_{\rm rc}$ are the central density, velocity
dispersion and relaxation time, respectively) form an attractor, which is
characterised by two large and
several very small loops. For both models (gaseous and Monte--Carlo) the
attractors projected on $v_{\rm c}$ plane are remarkably similar in
the range of
central density, $\zeta$ and as well in the shape and size of large loops.
This suggests that despite the strong stochasticity of $N$--body systems the
underlying physics of gravothermal oscillations for continuum models is
at least partially fulfilled.
It is worth to note that there are visible several short period and low
amplitude oscillations on Figure 8. These oscillations are probably
connected with stronger binary activities (significant energy generation)
and they are not mainly driven by the temperature inversion the core. So,
the picture that only part of oscillations are truly gravothermal
(McMillan \& Engle 1996) is further supported by the Monte--Carlo
simulations. In Figure 9 is shown the evolution of number of particles
in the core for $N=100000$. The number of particles at the phases of
maximum contraction is around $20$, while for the phases of maximum
expansion is around $1000$. This result is in excellent agreement with
$N$--body calculations (Makino 1996ab) and as well with gas and
Fokker--Planck calculations (Heggie \& Ramamani 1989, Breeden \et 1994).
This further strengthen the fact that gravothermal oscillations observed
in the Monte--Carlo simulations are practically undistinguished from that
for direct $N$--body simulations.

From the data shown it is concluded that oscillations are indeed present
in Monte--Carlo simulations with $N \geq 32000$ particles and possible with
$N = 10000$ particles. This is the first unambiguous detection (Giersz
1996) of gravothermal oscillations in Monte--Carlo simulations. This
finally closed the list of methods used to investigate evolution of
large collisional systems for which theoretical prediction of
gravothermal oscillations was confirmed.

Finally, a few words about the efficiency of the new code. The
calculations of $N=2000$, $4096$, $10000$ and $32000$ and $1000000$
models were conducted on PC--Pentium 166MHz and took about $2$, $6$, $20$,
$130$
and $2500$ hours, respectively. Taking into account the fact that $N$--body
simulations for $N=32768$, up to the core bounce, performed by
Makino (1996ab) took $3$ months on $1/4$ GRAPE--4 (Teraflop special--purpose
hardware) it can be easily shown that Monte--Carlo simulations are at least
$10^5$ times faster than $N$--body ones.
The theory for Monte--Carlo models predicts a linear increase of computing
time with $N$. This is
connected with the fact that the potential computation is the most time
consuming part of the code. In the Monte--Carlo model due to spherical
symmetry of the system the potential computation  is proportional to
$N$. However the real calculations show a steeper dependence on  $N$
than theory predicts. It seems that this is connected with the fact that
larger systems spend more time in phases  of high central density
(gravothermal oscillations). They
have  more density peaks, which imply smaller time steps in order to
properly resolve their evolution.

The high speed of the new Monte--Carlo code makes possible to run several
different models with modest $N$ to improve statistical quality of the
data and run individual models with $N$ as large as $100000$. This is a
first step in a direction to simulate real large $N$--body systems.

\section{CONCLUSIONS AND FUTURE DEVELOPMENTS OF THE NEW MONTE--CARLO CODE}
\medskip

A successful revision of Stod{\'o}{\l}kiewicz's Monte--Carlo code was
presented. The updated method treats each {\it superstar} as a single
star and follows the evolution and motion of all individual stellar
objects. This improvement was possible thanks to the recent developments
in computer hardware and computer speed. Two essential changes was added
to the original Monte--Carlo code. Firstly, the procedure which deal with
problems of radial velocity determination after the system rearrangement
(changes of mechanical energy of the stars due to changes of mass
distribution) was slightly changed. This assures better energy
conservation. Secondly, the new procedure which deals with star escapers
was added. This practically resolves the problem with too high escape
rate observed in Monte--Carlo simulations. The Monte--Carlo scheme
presented here (as previous Monte--Carlo schemes) takes full advantage of
the undisputed physical knowledge on the secular evolution of (spherical)
star clusters as inferred from continuum model simulations. Additionally it
describes in a proper way the graininess of the gravitational field and the
stochasticity of the real $N$--body systems. This does not include any
additional physical approximations or assumptions which are common in
Fokker--Planck and gas models (e.g. conductivity or isotropic distribution
function for field stars). From that respect Monte--Carlo scheme can be
regarded as a method which lies in the middle between direct $N$--body and
Fokker--Planck models and combines most advantages of the
both methods.

The first calculations for
equal--mass $N$--body systems with three--body energy generation
according to Spitzer's formulae show good agreement with direct
$N$--body calculations for $N=2000$, $4096$ and $10000$ particles. The
density, velocity, mass distributions, energy generation,
number of binaries etc. follow the $N$--body results. Only the number of
escapers is slightly too high compared to $N$--body results (but this can
be resolved by the time--dependent shift of the escape rate) and
there is no level off anisotropy for advanced post--collapse evolution of
Monte--Carlo models as is seen in $N$--body simulations for $N \leq 2000$.
For simulations with $N > 10000$ gravothermal 
oscillations are clearly visible. This is the first
unambiguous detection of gravothermal oscillations in Monte--Carlo
simulations. Moreover, this is a first unambiguous detection of
gravothermal oscillations for stochastic $N$--body system with $N$ as
large as $100000$.

The speed of the new code makes it possible to run individual models with
$N$ as large as $100000$ and also enables, in an unambiguous way, the
inclusion of several different physical processes which operate during
different stages of evolution of real globular clusters.

The new Monte--Carlo code described in this paper is seen as a first step
towards realistic models of globular clusters. Several important physical
processes have to be included to make the simulations of the stellar
systems more realistic. The final code will contain the following
physical processes: {\bf (1)} formation of binaries due to dynamical and
tidal interactions, {\bf (2)} primordial binaries, {\bf (3)} stellar
evolution,
{\bf (4)} tidal field of Galaxy and tidal shocks connected with crossing
the galactic plane and with large molecular clouds, {\bf (5)} collisions
between stars, {\bf (6)} interactions between binaries and stars and
between binaries themselves, improving the presently used scattering
cross-sections for binary hardening.

In the first stage all processes connected with interactions between
objects were modelled using analytical cross sections available in the
literature. This allowed the code to be tested, and made possible
comparison with continuum models.

In the next stage interactions between groups of three and four stars
will be modelled by numerical integrations of their orbits (the first
attempts are tested now). If during the integration the distance between
two or more stars becomes smaller than the sum of their radii then a
physical collision takes place. This more realistic approach ensures
that processes of energy generation  (the most important factor in the
dynamical evolution of globular clusters) will be modelled more closely.

The final stage will be the inclusion of detailed 3--D hydrodynamical
modelling of collisions between stars. This will be done by use of Smooth
Particle Hydrodynamics (SPH) for a limited number of particles per star
(a few hundred).  This will allow close comparison
between numerical models and observations of real globular clusters. I
refer here to observations of various, peculiar objects like blue
stragglers and milliseconds pulsars, which can be formed during
collisions and encounters between stars.

\bigskip
\bigskip

{\parindent=0pt
{\bf Acknowledgments}
I would like to thank Douglas C. Heggie and Rainer Spurzem for
stimulating discussions, comments and suggestions to a draft version
of this paper. I also thank Douglas C. Heggie, who made
the $N$--body results for $N = 4096$ particles available. This work was
supported in part by the Polish National
Committee for Scientific Research under grant 2--P304--009-06.}

\bigskip
\bigskip
\bigskip

\bsp

\end{document}